\documentclass[universe,communication,accept,pdftex,moreauthors]{Definitions/mdpi}

\firstpage{1} 
\pubvolume{1}
\issuenum{1}
\articlenumber{0}
\pubyear{2025}
\copyrightyear{2025}
\externaleditor{Firstname Lastname} 
\datereceived{25 July 2025} 
\daterevised{29 August 2025} 
\dateaccepted{30 August 2025} 
\datepublished{ } 
\hreflink{https://doi.org/} 

\newcommand{\aap}{Astron. Astrophys.}

\newcommand{\apj}{Astrophys. J.}
\newcommand{\apjs}{Astrophys. J. Suppl. Ser.}
\newcommand{\apjl}{Astrophys. J. Lett.}
\newcommand{\mnras}{Mon. Not. R. Astron. Soc.}

\newcommand{\pasp}{Publ. Astron. Soc. Pac.}

\newcommand{\araa}{Annu. Rev. Astron. Astrophys.}

\newcommand{\ssr}{Space Sci. Rev.}

\Title{What Do Radio Emission Constraints Tell Us About Little Red Dots as Tidal Disruption Events?}
\TitleCitation{What Do Radio Emission Constraints Tell Us About Little Red Dots as Tidal Disruption Events?}

\Author{Krisztina Perger
 $^{1,2,*}$\orcidA{}, Judit Fogasy $^{1,2}$\orcidB{} and Sándor Frey $^{1,2,3}$\orcidC{}}
\AuthorNames{Krisztina Perger, Judit Fogasy and S\'andor Frey}
\AuthorCitation{Perger, K.; Fogasy, J.; Frey, S.}
\address{%
$^{1}$ \quad Konkoly Observatory, HUN-REN Research Centre for Astronomy and Earth Sciences, Konkoly Thege Mikl\'{o}s \'{u}t 15-17, 1121 Budapest, Hungary; fogasy.judit@csfk.org (J.F.); frey.sandor@csfk.org (S.F.)\\
$^{2}$ \quad CSFK, MTA Centre of Excellence, Konkoly Thege Mikl\'{o}s \'{u}t 15-17, 1121 Budapest, Hungary\\
$^{3}$ \quad Institute of Physics and Astronomy, ELTE E\"otv\"os Lor\'and University, P\'azm\'any P\'eter s\'et\'any 1/A, 1117~Budapest, Hungary\\
}
\corres{Correspondence: perger.krisztina@csfk.org}

\abstract{The real nature of little red dots (LRDs), a class of very compact galaxies in the early Universe recently discovered by the James Webb Space Telescope, is still poorly understood. The most popular theories competing to interpret the phenomena include active galactic nuclei and enhanced star formation in dusty galaxies. To date, however, neither model gives a completely satisfactory explanation to the population as a whole; thus, alternative theories have arisen, including tidal disruption events (TDEs). By considering observational constraints on the radio emission of LRDs, we discuss whether TDEs are adequate alternatives solving these high-redshift enigmas. We utilise radio flux density upper limits from LRD stacking analyses, TDE peak radio luminosities, and volumetric density estimates. We find that the characteristic values of flux densities and luminosities allow radio-quiet TDEs as the underlying process of LRDs in any case, while the less common radio-loud TDEs are compatible with the model
under special constraints only. Considering other factors, such as volumetric density estimates, delayed and long-term radio flares of TDEs, and cosmological time dilation, TDEs appear to be a plausible explanation for LRDs from the radio point of view.}


\begin{document}
\section{Introduction}

Recent James Webb Space Telescope (JWST) photometric and spectroscopic observations revealed the presence of a new population of high-redshift ($z\ge4$) objects~\cite{2023ApJ...952..142F,2023Natur.616..266L,2024ApJ...963..128B,2024ApJ...964...39G,2025ApJ...986..126K,2024ApJ...963..129M,2025ApJ...978...92L}. These compact sources have a red continuum in the rest-frame optical waveband, show excess ultraviolet (UV) emission, and exhibit broad-line emission features in their spectra~\cite{2023ApJ...954L...4K,2024ApJ...964...39G,2024ApJ...963..129M}. Different theories were proposed to explain the nature of these so-called little red dots (LRDs), including highly obscured active galactic nuclei (AGN)~\cite{2025ApJ...985..169D,2024ApJ...964...39G,2025ApJ...986..126K}, enhanced star formation in dusty galaxies~\cite{2024ApJ...977L..13B,2024ApJ...968....4P,2024ApJ...968...34W,2025MNRAS.537.3453B}, or the combination of these two processes~\cite{2023ApJ...950L...5Y,2024ApJ...975..178K,2024ApJ...968....4P,2025MNRAS.537.3453B}.

Interestingly, the radio and X-ray properties of LRDs do not follow those of `regular' high-redshift AGN~\cite{2024ApJ...969L..18A,2024ApJ...974L..26Y,2025ApJ...986..126K,2024arXiv240610341A,2025A&A...693L...2P,2024arXiv241204224M}. To date, only one LRD was found to have a radio counterpart~\cite{2025ApJ...986..130G}, while the rest remain non-detected in the radio, even after stacking radio maps at the positions of known LRDs ~\cite{2024arXiv240610341A,2024arXiv241204224M,2025A&A...693L...2P}. Considering both radio-quiet and radio-loud scenarios, flux density detection thresholds of 0.5--2
~$\upmu$Jy were estimated, which could be uncovered via $10-100$~h long observations with upcoming sensitive state-of-the-art radio instruments, such as the Next-Generation Very Large Array and the Square Kilometre Array~\cite{2025A&A...694L..14L}. Similarly, only a handful X-ray-detected LRDs were found~\cite{2025ApJ...986..126K,2025MNRAS.538.1921M}, and the majority of the population remains silent, with only upper limits found on the X-ray fluxes via stacking~\cite{2024ApJ...969L..18A,2025ApJ...986..126K,2024ApJ...974L..26Y}. Moreover, LRDs remained undetected in the far-infrared regime~\cite{2024ApJ...968....4P,2024ApJ...968...34W,2025ApJ...978...92L}, with the absence of hot dust emission expected from an AGN torus~\cite{2024ApJ...968...34W, 2025ApJ...984..121W}. The extremely small fraction of radio- and X-ray-detected sources, the non-detection in the far infrared waveband, the missing torus-related emission, and the apparent dichotomy in the rest-frame colours and spectral properties of the population~\cite{2024ApJ...968...38K,2025arXiv250604350Z} raise the question of whether the LRDs are an extremely special case of active galaxies~\cite{2025A&A...693L...2P,2025MNRAS.539.2910P}, or whether the phenomenon might not be satisfactorily explained by the AGN model~\cite{2024ApJ...975L...4C,2024arXiv240704777K,2025arXiv250604004D}. 

Recently, an alternative model was suggested, explaining the observational properties of LRDs with tidal disruption events (TDEs) in dense clusters~\cite{2025ApJ...984L..55B}. In this model, it is theorised that black hole seeds forming through runaway collapse in dense clusters with a TDE rate of $10^{-4}$ per year would explain the compactness, UV emission, broad H$\alpha$ emission, and X-ray weakness of LRDs.

Tidal disruption events (TDEs) are rare transient phenomena, theorised in the early 1970s. 
TDEs occur when a star's orbit passes close enough to a massive black hole (MBH, $10^5-10^7$~M$_\odot$)~\cite{2017MNRAS.471.1694W,2020ApJ...904...73R}, causing the star to be pulled apart by the MBH's tidal forces, and a tidal stream of material is formed around the black hole. A portion of this material is captured by the MBH, forming an accretion disc, while producing luminous electromagnetic flares, which can last from weeks to years. To date, $\sim$150 TDEs and candidates have been identified~\cite{2023PASP..135c4101G}.

Approximately half of the TDEs were detected in one or more radio bands~\cite{2020SSRv..216...81A,2025ApJS..278...36G}, either close to the discovery date or during follow-up observations. In several cases, TDEs are accompanied by late-time radio flare-ups~\cite{2024ApJ...974..241A,2024ApJ...971..185C}, even up to thousands of days after discovery~\cite{2021ApJ...920L...5H,2024ApJ...971..185C}, with around $40\%$ of optical TDEs showing delayed activity in the radio~\cite{2024ApJ...971..185C}. Due to the inherently diverse radio properties of TDEs, various mechanisms were proposed and tested to explain the intrinsic radio emission, including external shocks with both relativistic~\cite{2011Natur.476..421B,2016ApJ...829...19V} and non-relativistic~\cite{2016ApJ...819L..25A,2023MNRAS.522.5084G} jets and internal~\cite{2018ApJ...856....1P,2022ApJ...933..176S} and external~\cite{2016ApJ...827..127K,2019MNRAS.487.4083Y} shock mechanisms due to interactions with or collisions within the created debris stream. Out of the several dozen radio-observed TDEs, only a few percent were found to be particularly bright in the radio wavebands, up to years or even decades after the initial disruption event~\cite{2016ApJ...819L..25A,2018Sci...361..482M,2022ApJ...925..220R}. These sources, labeled as radio-loud (RL) TDEs, are believed to host energetic radio jets at a small inclination angle with respect to the line of sight of the observer, releasing their energy with a specific luminosity of $\nu L_\nu>10^{40}$~erg~s$^{-1}$ (here, $\nu$ denotes the observing frequency). The remaining sources that emit their radio power below this limit are classified as radio-quiet (RQ) and often are not detected; alternatively, they are detected with a delay after the higher-energy observations~\cite{2017ApJ...844...46B,2021ApJ...920L...5H,2024ApJ...971..185C}.  RQ TDEs are believed to be either off-axis synchrotron jets dimmed via Doppler deboosting or are theorised to be slow outflows interacting with the surrounding circumnuclear medium, accelerating the free electron content~\cite{2020SSRv..216...81A,2021ARA&A..59...21G}. RL TDEs are usually discovered via their $\gamma$-ray outbursts and are also addressed as radio TDEs, while RQ TDEs are found by their initial detection in the X-ray, optical or UV bands~\cite{2020SSRv..216...81A,2020SSRv..216...85S,2020SSRv..216..124V}. As the optical--X-ray spectral energy distribution of RQ TDEs can be characterised by a black-body function peaking in the UV or the soft X-ray regimes~\cite{2009ApJ...698.1367G,2020SSRv..216..124V}, they are also referred to as thermal TDEs. The occurrence rate of radio-emitting TDEs with powerful radio jets was estimated to be $3\cdot10^{-10}$~yr$^{-1}$~galaxy$^{-1}$~\cite{2015MNRAS.452.4297B}, while for thermal TDEs, this rate was found to be $5\cdot10^{-5}$~yr$^{-1}$~galaxy$^{-1}$~\cite{2020SSRv..216...81A}. Other works reported an overall TDE rate, i.e., a value for the ensemble of both thermal and radio TDEs, within the order of magnitude of $\sim10^{-4}$~yr$^{-1}$~galaxy$^{-1}$ ~\cite{2016MNRAS.455..859S,2018ApJ...852...72V}.

In this paper, we explore the possibility of TDEs as an explanation for LRDs, based on the extreme quiescence of the latter in the radio regime.

\section{Methodology}

To test the possibility of TDEs as the underlying physical processes of LRDs, we calculated upper limits to the rest-frame specific luminosities of LRDs based on their radio flux density upper limits, following the equation  
\begin{linenomath}
\begin{equation}
\nu L_\nu=4\pi\,D_\mathrm{L}^2\,S\,\nu\,(1+z)^{-\alpha-1},
\end{equation}
\end{linenomath}
where $S$ is the flux density upper limit from the respective stacking analysis, $ D_\mathrm{L}$ is the luminosity distance at redshift $z$, and $\alpha$ is the power-law spectral index, following the $S_\nu\sim\nu^\alpha$ convention. For the calculations, we applied a standard flat $\Lambda$CDM cosmological model with $H_0 = 70$~km~s$^{-1}$~Mpc$^{-1}$, $\Omega_\mathrm{M}=0.3$, and $\Omega_\Lambda=0.7$. 

For the analysis described below, we assumed that the whole LRD population is caused by TDEs, as proposed by~\cite{2025ApJ...984L..55B}. The properties of the representative TDEs selected for the analysis are listed in Table~\ref{tab:tdes}. Peak radio luminosity values were adopted from~\cite{2020SSRv..216...81A}.

\begin{table}[H]
\caption{TDEs selected for comparison from the sample presented by~\cite{2020SSRv..216...81A}}.\label{tab:tdes}
\setlength{\tabcolsep}{6.6mm} 
\begin{tabular}{cccc}
\toprule
\textbf{Name}	& \textbf{Type} & \textbf{Redshift} 	& \textbf{Luminosity (erg~s$^{-1}$})\\
\midrule
XMMSL1~J0740$-$85	& RQ	& 0.0173	& $1\cdot10^{37}$	\\
ASASSN-14li	  & RQ	& 0.0206	& $9\cdot10^{37}$	\\
Sw~J1112$-$82	& RL	& 0.8901	& $3\cdot10^{40}$	\\
Sw~J2058+05	  & RL	& 1.1853	& $8\cdot10^{41}$	\\
\bottomrule
\end{tabular}
\end{table}

\subsection{Rest-Frame Luminosity Estimation of LRDs}\label{sec:methodL}
Rest-frame luminosities of LRDs were calculated using the parameter space of various flux densities, redshifts, and spectral indices (hereafter Method L). For flux densities, $3$-GHz upper limits ($S\le 10~\upmu$Jy and $S\le 0.5~\upmu$Jy) were used, while redshift ranges ($2.4\le z\le 11.4$ and $2.26\le z\le 10.26$) were defined by adopting the values of the known LRDs in the respective stacking analyses at $3$~GHz~\cite{2025A&A...693L...2P,2024arXiv241204224M}. Based on the stacking analysis on radio maps from the Very Large Array Sky Survey (VLASS) \citep{2020RNAAS...4..175G,2021ApJS..255...30G} of $919$ objects, a $3\sigma$ upper limit of $\sim$10~$\upmu$Jy was found for the flux density of LRDs~\cite{2025A&A...693L...2P}. Alternatively, we used $3\sigma$ upper limits of $0.5~\upmu$Jy determined for $22$ broad-line objects based on deep radio observations in the GOODS-N field~\cite{2024arXiv241204224M}. We note that a similar flux density upper limit was found by stacking of maps of $434$ LRDs selected from the COSMOS field Very Large Array observations~\cite{2024arXiv240610341A}. For spectral indices, we considered various values ($-2.05\le \alpha \le 2.55$) found for TDEs in the literature~\cite{2021ApJ...920L...5H,2021NatAs...5..491H,2022ApJ...925..220R,2023MNRAS.520.2417W,2024ApJ...973..104D,2024ApJ...974..241A,2025ApJ...983...29H}. We compare the estimated luminosity ranges of LRDs with values observed in two RQ and two RL TDEs (Table~\ref{tab:tdes}) in  Section~\ref{sec:results_lumiosity}.

\subsection{Flux Density Estimation of `High-Redshift' TDEs}\label{sec:methodS}
In the other way around (hereafter Method S), $3$-GHz flux densities were calculated for a RQ (thermal) and a RL (radio) TDE, Sw~J2058.4+05 and XMMSL1~J0740$-$85, respectively, for the whole redshift range ($2.26\le z\le 11.4$) covered in the two stacking analyses~\cite{2025A&A...693L...2P,2024ApJ...963..129M}, using the following relation:
\begin{linenomath}
\begin{equation}
S=\frac{\nu L_\nu}{4\pi\,D_\mathrm{L}^2\,\nu\,(1+z)^{-\alpha-1}},
\end{equation}
\end{linenomath}
where $\nu L_\nu$ is the specific luminosity of the respective TDE. For the estimation, rest-frame peak radio luminosities were used (listed in Table~\ref{tab:tdes}), considering a wide range of spectral indices of known TDEs ($-2.05\le \alpha \le 2.55$) from the literature~\cite{2021ApJ...920L...5H,2021NatAs...5..491H,2022ApJ...925..220R,2023MNRAS.520.2417W,2024ApJ...973..104D,2024ApJ...974..241A,2025ApJ...983...29H}. The range of possible flux densities was compared to the upper limits determined for LRDs~\cite{2024arXiv241204224M,2025A&A...693L...2P}.

\subsection{Volumetric Number Density Estimation}
Redshift-dependent TDE rates $\mathrm{TDE_1}(z)$ and $\mathrm{TDE_2}(z)$ were estimated following two separate relations. Values for $\mathrm{TDE_1}(z)$ were calculated using the TDE rate--redshift relation from~\cite{2015ApJ...812...33S} scaled by a factor of  $4\cdot10^{-6}$~Mpc$^{-3}$~\cite{2009ApJ...697L..77R}. Alternatively, $\mathrm{TDE_2}(z)$ values were calculated, combining the $\sim(1+z)^{-2.8}$ density--redshift relation for galaxies~\cite{2004ApJ...606L..25B} with rates of $2\cdot10^{-4}$~\cite{2016MNRAS.455..859S}, $5\cdot10^{-5}$~\cite{2020SSRv..216...81A}, and $3\cdot10^{-10}$~galaxy$^{-1}$~yr$^{-1}$~\cite{2015MNRAS.452.4297B}, determined for a mixed sample, RQ, and RL TDEs, respectively. The typical rate of TDEs varies between $10^{-5}$ and $10^{-9}$~Mpc$^{-3}$~yr$^{-1}$ in the redshift range $0.35\le z\le 12$. 
\section{Results and Discussion}

\subsection{Luminosities of LRDs}\label{sec:results_lumiosity}

\begin{figure}[H]
  \includegraphics[width=0.75\linewidth]{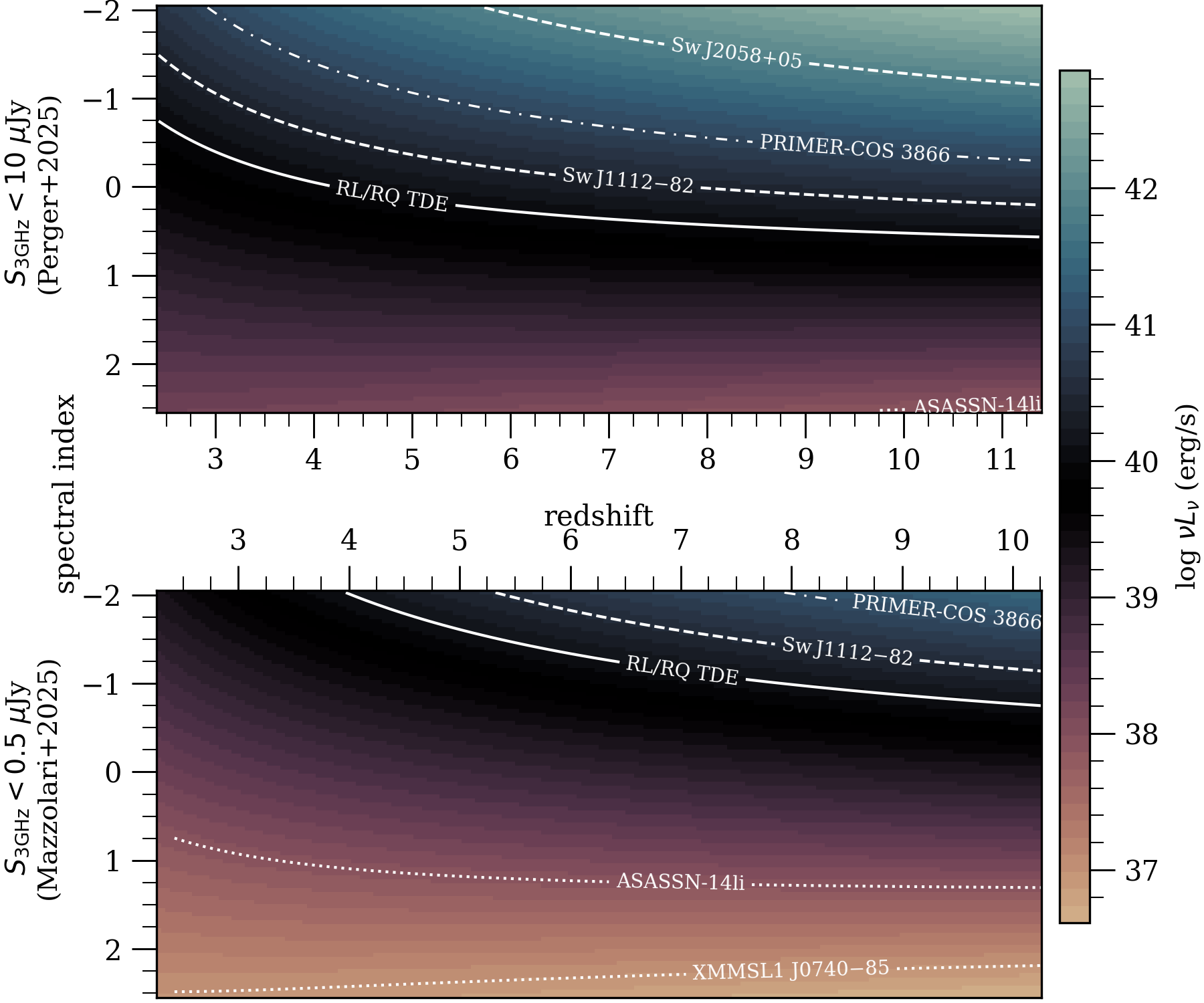}
  \caption{Parameter space allowed by the $3$-GHz flux density upper limits~\cite{2025A&A...693L...2P,2024arXiv241204224M} on LRDs. Possible radio spectral index ($\alpha$) values are shown as a function of redshift ($z$), while the colour scale indicates the logarithm of specific luminosity ($\nu L_\nu$) values. Contours of peak radio luminosities~\cite{2020SSRv..216...81A} of the four selected representative TDEs (Table~\ref{tab:tdes}) are denoted with dashed and dotted contours for the RL and RQ sources, respectively. The RL/RQ threshold ($\nu L_\nu>10^{40}$~erg~s$^{-1}$) dividing the radio and thermal TDE populations is shown with a solid line. For comparison, the luminosity~\cite{2025ApJ...986..130G} of the only LRD detected to date in radio bands (PRIMER-COS~3866) is indicated with a dash-dotted line.}
  \label{fig:lumi}
\end{figure}

\textls[-15]{The luminosity ranges determined for the LRD upper limits based on the parameter space of redshifts and spectral indices of known TDEs (Method L, Section~\ref{sec:methodL}) are illustrated in Figure~\ref{fig:lumi}. The values calculated at redshifts $2.4\le z\le 11.4$ and $2.26\le z\le 10.26$ are in agreement with those found in low-redshift known TDEs. The $S<10~\upmu$Jy flux density upper limit~\cite{2025A&A...693L...2P} found from stacking the $3$~GHz radio maps of $919$ LRDs allows both RL and RQ TDEs (Figure~\ref{fig:lumi}, upper panel). On the other hand, the more strict constraint of $S<0.5~\upmu$Jy~\cite{2024arXiv241204224M} found by stacking the more sensitive radio images of a smaller sample of LRDs only allows RL TDEs at higher redshifts, while the majority of the parameter space only allows RQ TDEs as the origin of the LRD population (Figure~\ref{fig:lumi}, lower panel). The only radio-detected LRD to date, PRIMER-COS~3866 ($z=4.66$), has a luminosity of $\nu L_\nu=9\cdot10^{40}$~erg~s$^{-1}$~\cite{2025ApJ...986..130G}. This would place PRIMER-COS~3866 among the more powerful RL TDEs. }

\subsection{Flux Densities of TDEs}

\begin{figure}[H]
  \includegraphics[width=0.75\linewidth]{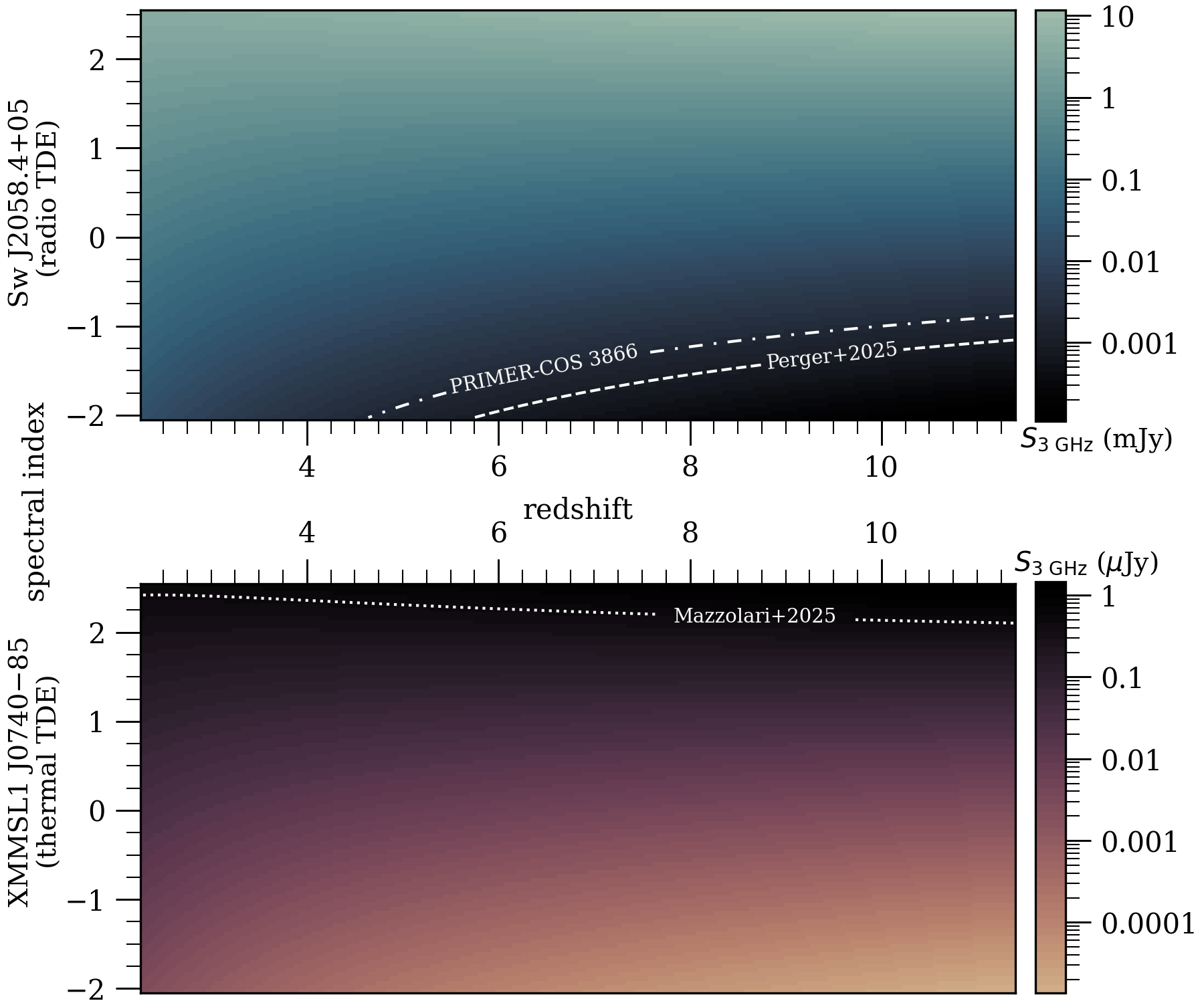}
  \caption{Parameter space allowed by the luminosities of RL and RQ TDEs, Sw~J2058.4+05 and XMMSL1~J0740$-$85, respectively. Possible radio spectral index ($\alpha$) values are shown as a function of redshift ($z$), while the colour scale indicates the 3-GHz radio flux density ($S$) values. The upper limits on the  radio flux densities found in the stacking studies~\cite{2025A&A...693L...2P,2024arXiv241204224M} are denoted with dashed and dotted contours, respectively. For comparison, the flux density~\cite{2025ApJ...986..130G} of the only LRD detected to date in radio bands (PRIMER-COS~3866) is indicated with a dash-dotted line.} \label{fig:fluxd}
\end{figure}

The flux densities estimated from peak radio luminosities of known TDEs (Method S, Section~\ref{sec:methodS}) for the total redshift range ($2.26\le z\le 11.4$) of LRDs are shown in Figure~\ref{fig:fluxd}. The flux density ranges allowed by the parameter space of redshift, spectral index, and TDE \textbf{luminosity} suggest that LRDs are consistent with an underlying process from thermal/RQ TDEs, considering both~\cite{2025A&A...693L...2P} and~\cite{2024arXiv241204224M}. On the other hand, energetic radio TDEs such as  Sw~J2058.4+05 ($\nu L_\nu\sim10^{41}$~erg~s$^{-1}$) are only permitted at redshifts $z\ge6$ with a very steep spectrum, $-2\le\alpha\le-1$. These constraints permit objects with similar spectra as the RQ TDEs with delayed radio emission, like ASASSN-15oi ($\nu L_\nu = 7\cdot10^{38}$~erg~s$^{-1}$)~\cite{2021NatAs...5..491H} and  AT2019dsg ($\nu L_\nu = 4\cdot10^{38}$~erg~s$^{-1}$)~\cite{2021NatAs...5..510S,2022ApJ...927...74M}. To date, only one object, PRIMER-COS~3866 ($\nu L_\nu=9\cdot10^{40}$~erg~s$^{-1}$)~\cite{2025ApJ...986..130G}, has been detected in radio wavebands out of $\sim900$ known LRDs. Thus, the radio-detected fraction of LRDs is estimated to be $10^{-3}$. Considering that RQ sources dominate the characteristic properties of LRDs, and that the dominant population of TDEs is also radio-quiet, even when detected in radio bands, the connection between the two phenomena cannot be ruled out. 

\unskip

\subsection{Number Densities}
\begin{figure}[H]
    \includegraphics[width=0.85\linewidth]{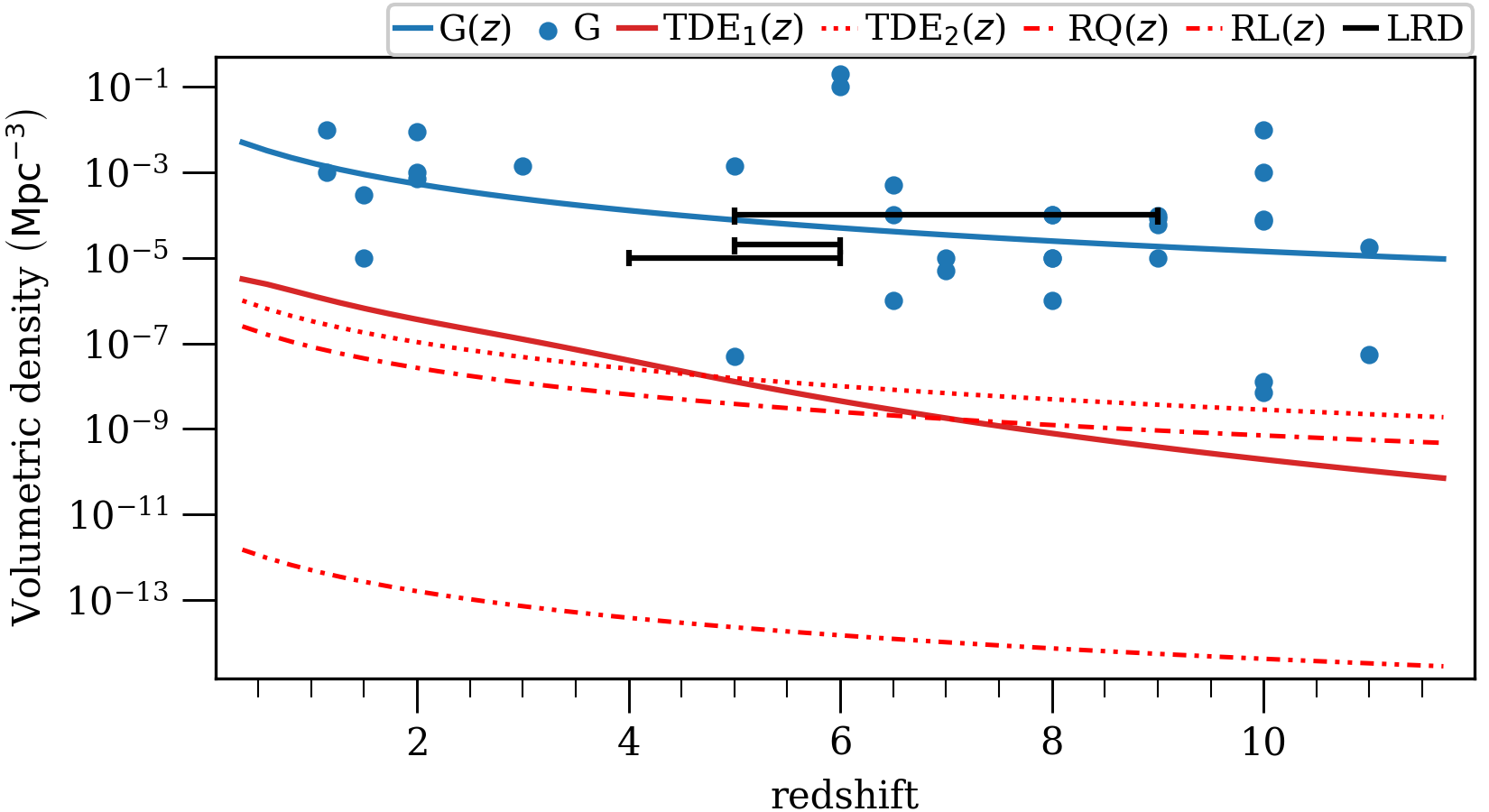}
    \caption{Volumetric density of galaxies (red), little red dots (black), and tidal disruption event occurrence rates (blue). The solid blue line denotes the redshift dependence of galaxies following the relation of $\sim$$(1+z)^{-2.8}$~\cite{2004ApJ...606L..25B} with a scaling factor of $1.2\cdot10^{-2}$~\cite{2023A&A...670A..77M}. Independent volumetric density estimations from the literature are denoted with blue bullet points~\cite{2004MNRAS.355..374B,2013MNRAS.429.2098W,2015ApJ...803...34B,2016MNRAS.459.3812M,2016MNRAS.461.1112O,2020MNRAS.493.2059B,2023A&A...672A..71M,2024ApJ...964..161M,2024ApJ...969L...2F,2024ApJ...973...23F,2025A&A...698A.103B,2025ApJ...978...89H,2025ApJ...985...80R}. 
    For densities given within redshift ranges, mean values were used for the plot. The thick solid red line denotes the rate--redshift relation of TDEs~\cite{2015ApJ...812...33S}, scaled by a factor of  $4\cdot10^{-6}$~Mpc$^{-3}$~\cite{2009ApJ...697L..77R}. The dotted, dash-dotted, and dash-dot-dotted lines denote TDE occurrence rates per volumetric density combining the galaxy volumetric relation of $\sim$$(1+z)^{-2.8}$~\cite{2004ApJ...606L..25B} and rates of $2\cdot10^{-4}$~\cite{2016MNRAS.455..859S}, $5\cdot10^{-5}$~\cite{2020SSRv..216...81A}, and $3\cdot10^{-10}$~galaxy$^{-1}$~yr$^{-1}$~\cite{2015MNRAS.452.4297B}, for all, RQ, and RL TDEs, respectively. 
    Volumetric densities of little red dots~\cite{2024ApJ...964...39G,2024ApJ...963..129M,2024ApJ...968....4P} are shown with closed black line segments.}
    \label{fig:rates}
\end{figure}

The relation between volumetric number densities of galaxies, LRDs, and TDE rates is illustrated in Figure~\ref{fig:rates}. The number density of LRDs  was found to be approximately constant in the range of $2\cdot 10^{-5}-10^{-4}$~Mpc$^{-3}$ at redshifts $z\ge4$~\cite{2024ApJ...964...39G,2024ApJ...963..129M,2024ApJ...968....4P}, while rapidly declining in the lower-redshift regime~\cite{2025ApJ...986..126K}. LRD densities are consistent with values found for regular galaxies~\cite{2013MNRAS.429.2098W,2015ApJ...803...34B,2020MNRAS.493.2059B,2023A&A...672A..71M,2025A&A...698A.103B,2025ApJ...978...89H,2025ApJ...985...80R} within the same redshift domain. Based on this, if LRDs are indeed associated with TDEs~\cite{2025ApJ...984L..55B}, the expected number of TDEs would be in the same order of magnitude as in `regular' galaxies. Assuming the most optimistic scenario, i.e., where the ratio of the TDE rate to the galaxy number density is approximately constant (within an order of magnitude,  $\mathrm{TDE_2}(z)$), the evolution of TDE rates with respect to the redshift (Figure~\ref{fig:rates}) would predict one TDE for $\sim$$10^3$~galaxy~yr$^{-1}$. Considering the number or LRDs ($919$) and their radio-detected fraction, we would expect radio LRDs in the same order of magnitude as the predicted TDE rate in the best-case estimation. Considering the similar occurrence rate, the long-term nature of the radio emission in TDEs lasting for as long as decades, and that the cosmological time dilation by the factor $(1+z)$ at the redshifts of LRDs would make the radio emission detectable up to $\sim$100 years in the observer's frame, a population of high-redshift tidal disruption events as the hidden process behind JWST little red dots could be consistent with their extremely radio-quiet nature.

It is important to note, as discussed by~\cite{2025ApJ...984L..55B}, that the observational properties of the known LRDs are not completely uniform. 
Thus, it is plausible that these high-redshift sources may reflect a diversity of astrophysical objects and processes rather than a single universal model, raising the possibility of distinct subsets of the LRD population.

\section{Conclusions}

We found that the inferred luminosity values calculated from the flux density upper limits from radio stacking analyses focusing on little red dots~\cite{2025A&A...693L...2P,2024arXiv241204224M} are consistent with the observed values of known tidal disruption events. Considering the radio luminosity at peak value, the less strict constraints~\cite{2025A&A...693L...2P} allow the occurrence of powerful radio (RL) TDEs as an explanation for the generally extremely radio-quiet LRDs, while the luminosities estimated from the more stringent flux density limit of~\cite{2024arXiv241204224M} only allow thermal (RQ) TDEs to be those responsible for producing LRDs. Taking into account that TDEs tend to be radio-detected up to decades after the initial outburst, the cosmological time dilatation at the redshift range of LRDs, the order-of-magnitude agreement between the TDE occurrence rate of $10^{-3}$~yr$^{-1}$ and that of the radio LRDs, the low fraction of RL TDEs~\cite{2020SSRv..216...81A}, and the number of known LRDs~\cite{2025A&A...693L...2P}, we conclude that, based on the radio properties, tidal disruption events are possible explanations of little red dots. As LRDs were found not to be an entirely homogeneous population, either of the previously discussed models (AGN, dusty star-forming galaxies, or TDEs) could explain distinct subsets of these mysterious objects.

\vspace{6pt} 

\authorcontributions{Conceptualization, S.F.; methodology, K.P.; validation, J.F. and S.F.; formal analysis, K.P.; investigation, K.P.; data curation, K.P.; writing---original draft preparation, K.P.; writing---review and editing, J.F. and S.F.; visualization, K. P.; funding acquisition, J.F. and S.F. All authors have read and agreed to the published version of the manuscript.}

\funding{This research was funded by the Hungarian National Research, Development and Innovation Office (NKFIH), grant numbers OTKA K134213 and PD146947, and by the NKFIH excellence grant TKP2021-NKTA-64.}

\dataavailability{No new data were created in this study. Data sharing is not applicable to this article.} 

\conflictsofinterest{The authors declare no conflicts of interest.}

\abbreviations{Abbreviations}{
The following abbreviations are used in this manuscript:\\

\noindent 
\begin{tabular}{@{}ll}
AGN & active galactic nuclei\\
JWST & James Webb Space Telescope\\
LRD & little red dot\\
RL & radio-loud\\
RQ & radio-quiet\\
TDE & tidal disruption event\\
UV & ultraviolet\\
VLASS & Very Large Array Sky Survey
\end{tabular}
}
\begin{adjustwidth}{-\extralength}{0cm}

\reftitle{References}

\PublishersNote{}
\end{adjustwidth}
\end{document}